# Surveying students' understanding of quantum mechanics in one spatial dimension

Guangtian Zhu and Chandralekha Singh

Department of Physics and Astronomy, University of Pittsburgh, Pittsburgh, Philadelphia, 15260, USA

*Abstract*

We explore the difficulties that advanced undergraduate and graduate students have with non-relativistic quantum mechanics of a single particle in one spatial dimension. To investigate these difficulties we developed a conceptual survey and administered it to more than two hundred students at eleven institutions. The issues targeted in the survey include the set of possible wavefunctions, bound and scattering states, quantum measurement, expectation values, the role of the Hamiltonian, and the time-dependence of the wavefunction and expectation values. We find that undergraduate and graduate students have many common difficulties with these concepts and that research-based tutorials and peer-instruction tools can significantly reduce these difficulties. The findings also suggest that graduate quantum mechanics courses may not be effective at helping students develop a better conceptual understanding of these topics, partly because such courses mainly focus on quantitative assessments.

## I. INTRODUCTION

Learning quantum mechanics is challenging.[1-4] The concepts are not intuitive and is very different from the ones which students are used to from their previous courses and everyday experiences.[5] Moreover, a good understanding of the formalism of quantum mechanics requires a solid grasp of linear algebra, differential equations, and special functions. Despite the mathematical facility required to master quantum mechanics, the formalism has a coherent conceptual framework.[6-8]

For student learning to be meaningful, the goals of the course, the instructional design, and the assessment of learning should all be aligned.[9-11] Because students will focus on what is assessed, assessment should include an understanding of the conceptual framework and knowledge structure of quantum mechanics. Without a conceptual framework, students are unlikely to retain what they have learned when the course is over.

Multiple-choice conceptual surveys are useful tools for evaluating students' understanding of various topics. Such surveys are easy to administer and grade. The results are objective and amenable to statistical analysis so that different instructional methods and different student populations can be compared. The Force Concept Inventory is a conceptual multiple-choice survey which has helped instructors recognise that many introductory physics students do not develop a functional understanding of force concepts even if they can solve quantitative problems. Other conceptual surveys have been designed for many physics topics, including energy and momentum, rotational and rolling motion, circuits, electricity and magnetism, and Gauss's law.[12] These surveys reveal that students have many conceptual difficulties with classical physics. Research-based instructional strategies have been shown to significantly improve students' conceptual understanding of some of these topics.[10-11]

To explore the conceptual difficulties that undergraduate and graduate students have with quantum mechanics, we developed the Quantum Mechanics Survey (QMS), a 31-item multiple-choice test. The survey was developed by consulting with many quantum mechanics instructors about the goals of their undergraduate courses and the topics their students should have learned. We then iterated different versions of the open-ended and multiple-choice questions with a subset of these instructors during the development of the survey. To investigate students' difficulties with various concepts, we administered free-response and multiple-choice questions and conducted interviews with individual students using a think-aloud protocol. In this interview protocol, students were asked to talk aloud while they answered the questions so that the interviewer could record their thought processes. Individual interviews with students during the investigation of difficulties and the development of the survey were useful for an in-depth understanding of students' thought processes.

Undergraduate quantum mechanics is sometimes taught as a one semester course. Also, some instructors begin with two-state systems before covering quantum mechanics of a single particle in one dimension. Although such courses may help students develop a good grasp of quantum mechanics, all concepts covered in the survey may not be discussed in such courses. Our survey is not appropriate for such courses in which all relevant concepts are not covered.

**II. SURVEY DESIGN**

The survey focuses on assessing students' understanding of the conceptual framework of quantum mechanics of one particle in one spatial dimension rather than assessing their mathematical skills. Students can answer the survey questions without performing any complicated mathematics, although students need to understand the basics of linear algebra. Because the survey focuses on quantum systems in one dimension, the concept of orbital angular momentum is not included in the survey. We also did not include spin angular momentum and Dirac notation to ensure that it can be used after most junior/senior-level quantum mechanics courses regardless of textbook, institution, or instructor.

While designing the survey, we paid particular attention to reliability and validity.[13-14] Reliability refers to the degree of consistency between individual scores if someone immediately repeats the test; validity refers to the appropriateness of interpreting the test scores. To ensure that the survey is valid, we took into account the opinions of 12 instructors regarding the goals of junior/senior-level quantum mechanics courses and the concepts that their students should have learned.[15] We also surveyed faculty members who had taught a two semester upper-level undergraduate course about these issues at a 2002 Gordon Research Conference on quantum mechanics. We found many commonalities about what these instructors expected their students to have learned. In addition to using pen and paper (or online) surveys, we discussed these issues individually with several instructors at the University of Pittsburgh who have taught quantum mechanics at the junior-senior and/or graduate level.

The quantum mechanical models in the survey are all confined to one spatial dimension (1D), for example, the infinite/finite square well, the simple harmonic oscillator, and the free particle. The survey includes a wide range of topics such as the possible wavefunction, the expectation value of a physical observable and its time dependence, the role of the Hamiltonian, stationary and non-stationary states and issues related to their time development, and measurements.

Before developing the questions for the survey, we developed a test blueprint to provide a framework for deciding the desired test attributes. The specificity of the test plan helped us to determine the extent of content covered and the complexity of the questions. In developing good alternatives for the multiple-choice questions, we took advantage of prior work on student difficulties with quantum mechanics.[16-20] To investigate student difficulties further, we administered a set of free-response questions in which students had to provide their reasoning. The answers to these open-ended questions were summarized and categorized, which helped us develop alternatives for the questions in the survey based on common difficulties. The incorrect choices often had distracters which reflect students' common misconceptions to increase the discriminating properties of the questions. Having good distracters in the alternative choices is important so that the students do not select the correct answer for the wrong reason. Statistical analysis was conducted on the preliminary versions of the multiple-choice questions to help refine the questions further.

We interviewed individual students using a think-aloud protocol[21] to develop a better understanding of students' reasoning processes when they were answering the open-ended and multiple-choice questions. During the think-aloud interviews, some previously unnoticed difficulties and misconceptions were revealed. These common difficulties were incorporated into new versions of the written tests and ultimately into the multiple-choice questions in the survey. Four professors at the University of Pittsburgh reviewed different versions of the survey several times to examine its appropriateness and relevance for upper-level undergraduate quantum mechanics courses and to detect any ambiguities in item wording. Many professors from other universities also provided valuable comments and feedback to fine-tune the survey. Each question has one correct choice and four incorrect choices.[13]

Some of the questions were based on the research-based learning tools for quantum mechanics such as concept tests[22] and Quantum Interactive Learning Tutorials.[17] Most of the upper-level students enrolled in a two semester quantum mechanics course are able to complete the survey in one class period after all these topics are covered in class. Experience in introductory physics suggests that physics professors often take a significantly longer time to answer the questions in the Force Concept Inventory when they take it for the first time compared to students (most of whom finish it in less than 30 minutes both before and after instruction in relevant concepts).

**III. SURVEY RESULTS**

The survey was administered to 226 students from ten universities. Although ten universities were involved, 14 different classes were administered the survey because both the upper-level undergraduate and graduate classes took it at one institution for two consecutive years. Among the 226 students, 33 were first year graduate students enrolled in a two semester graduate quantum mechanics course. The survey was administered after the first semester. The other students were undergraduates who had taken at least a one-semester quantum mechanics course at the junior/senior level. All students completed the survey in one class period except those in a class where the instructor taught quantum mechanics in two back-to-back class periods. This instructor requested that his students be allowed to use both back-to-back class periods to complete the survey. Because there is no statistically significant difference between the scores of these students and those from other classes, we do not distinguish between these students. Two of the junior/senior classes where students were enrolled in a two semester course used research-based learning tools such as concept tests[22] and Quantum Interactive Learning Tutorials. The survey was given twice, once at the end of the first semester (28 students) and then again at the end of the second semester (26 students).

The average score on the survey for all 226 students regardless of instruction (including only the first score of students who took it twice) is 45%. The reliability coefficient $\alpha$ (which is a measure of the internal consistency of the test with a high $\alpha$ signifying that some students consistently perform well across various questions on the test while others perform poorly) is 0.91, which is quite good by the standards of test design.[13] The percentage of students who correctly answered each question is shown in Fig. 1 and ranged between 0.2 and 0.8. Most of the percentages were around 0.4. This range is consistent with our previous investigations of student difficulties. Figure 2 shows the item discrimination, which represents the ability of a question to distinguish between the high and low performing students in the overall survey. A measure of item discrimination is the point biserial discrimination coefficient,[13] which is the correlation between the score on a particular question for each student and the total test score minus the score on that question for each student. The point biserial discrimination coefficient ranged from approximately 0.3 to 0.6 with about 3/4 of the questions with point biserial discrimination coefficients higher than 0.4. The standards of test design[13] indicate that the survey questions have reasonably good item discrimination.

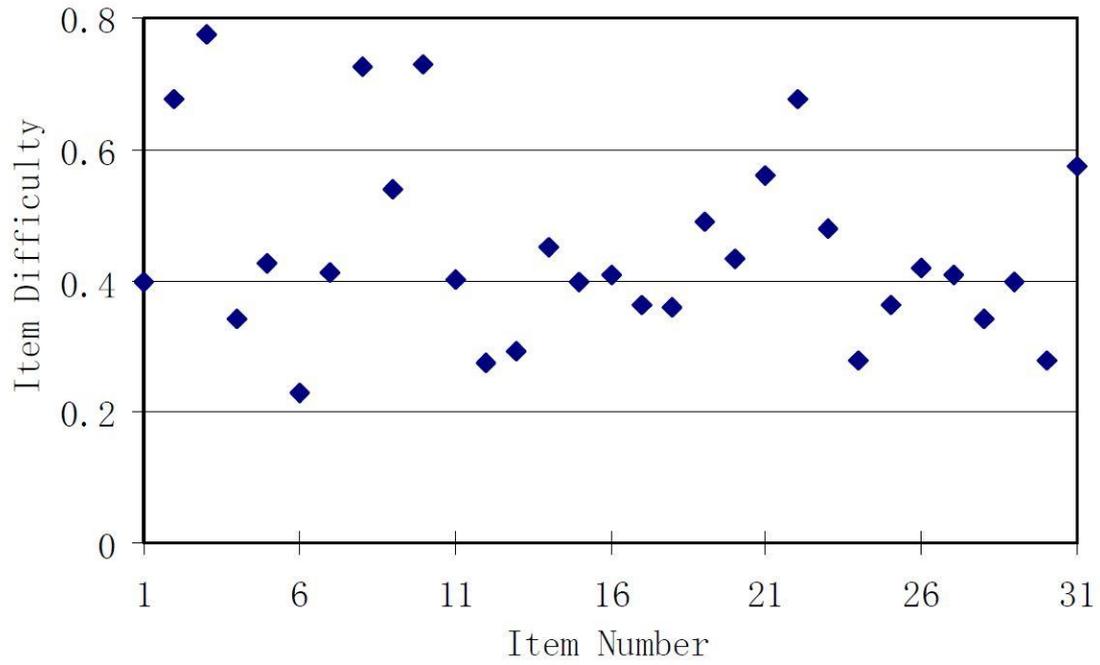

Fig 1. Item difficulty (fraction correct) for each item on the test for 226 students.

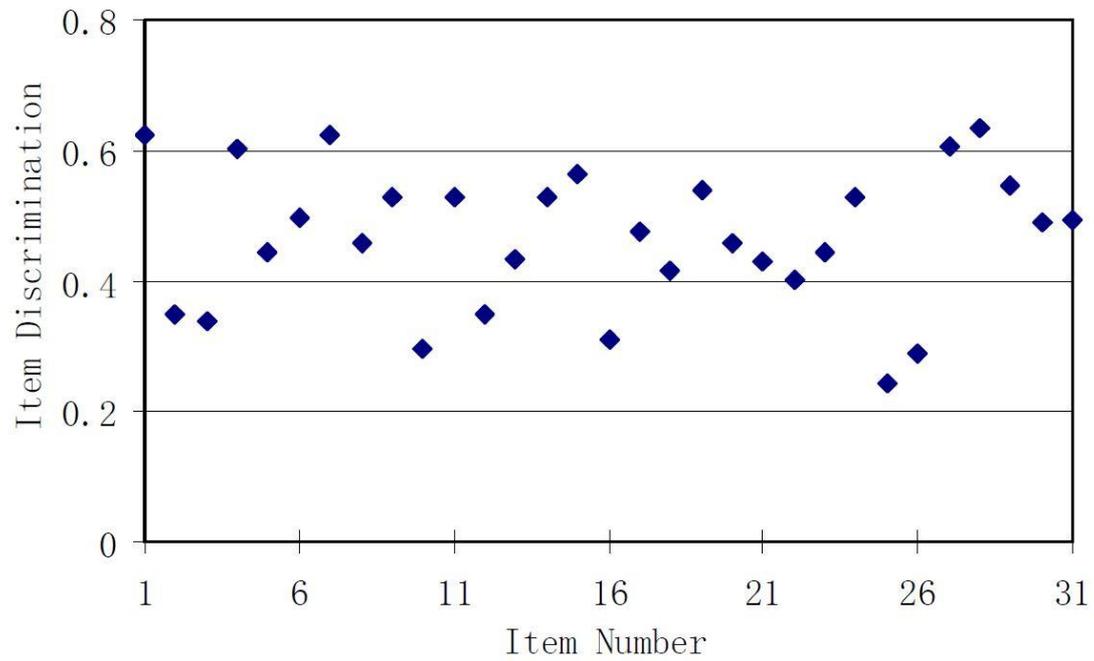

Fig 2. Item discrimination for each item on the test.

The average score for the upper-level undergraduate classes that used concept tests and Quantum Interactive Learning Tutorials during the semester was 71.5% at the end of the first semester (28 students) and 69.4% at the end of the second semester (26 students). The average performance of students who used the research-based active learning tools[17,22] did not deteriorate after a second semester in quantum mechanics. In classes that did not use the learning tools, the average score was 51.6% for the graduate course (33 students) and 39.0% for the undergraduate courses. Note that although students would score 20% on average if they answered all questions randomly given a five item multiple choices, experience with the Force Concept Inventory in introductory physics suggests that with good distracters students' performance can often be worse than random because they find the distracters attractive.[12] Our item analysis (to be discussed) suggests that students are not randomly guessing and are providing responses they think are reasonable. (In individual interviews students often claim that the alternative choices are the correct choices for those questions.)

Although the graduate student performance is low, discussions with two graduate quantum mechanics course instructors suggest that they expected their students to know all the survey content and perform well. After realizing that the graduate students had not done so, the graduate instructors agreed that many of the graduate students lacked conceptual understanding necessary for performing well on the survey even though they do well on the quantitative exams typically given in the graduate level courses. The poor performance of the graduate students suggests that they would develop a more robust knowledge structure if graduate quantum courses focused on both conceptual and quantitative problem solving (rather than only quantitative problem solving) by including conceptual problems in the assessment of student learning.

## IV. ITEM ANALYSIS

The survey is included in the supplementary material.[23] Table I shows a particular categorization of the questions in the survey based on the concepts. The table provides one of the possible ways to classify the questions. Our prior research shows that instructors categorize a given question in many different ways[15] so the categorization in Table I is only one of those which we found convenient. In the group "Other," Question 21 is about the uncertainty principle; Question 25 involves the concept of degeneracy in the context of a free particle; and Question 26 involves the Ehrenfest theorem, which states that the time dependence of the expectation value of a physical observable of a quantum system obeys the classical laws. In the following, we describe the common difficulties found by the survey in each of the categories.

| Concepts | Item Number |
|---|---|
| Possible wavefunctions | 1, 6, 14, 16, 30 |
| Bound/scattering states | 18, 19, 24, 27, 31 |
| Measurement | 5, 7, 8, 11, 13, 17, 20, 22, 28 |
| Expectation values | 9, 12, 25 |
| Time dependence of expectation values | 2, 10, 23, 26 |
| Stationary and non-stationary states | 3, 4, 6, 15, 20, 25, 28, 29 |
| Role of the Hamiltonian | 26, 27, 29 |
| Time dependence of wavefunction | 3, 4, 6, 15, 17, 22, 29 |
| Other | 21, 25, 26 |

Table I. A possible categorization of the QMS items and the question numbers belonging to each category

### A. The Possible Wavefunctions

Tables II to VIII show the percentages of students selecting the choices (a)–(e) on the problems in different categories, e.g., the possible wavefunctions, stationary states, etc. The correct responses are in boldface. In some columns the percentages of choices do not sum to 100% because some students left a question blank. A very common misconception about the possible wavefunctions is thinking that only certain forms of the wavefunctions are allowed. Students usually encounter the energy eigenstates (or position eigenstates) when they are learning quantum mechanics, so they often think that the possible wavefunctions must be energy eigenstates or eigenstates of an operator corresponding to a physical observable. A superposition of the energy eigenfunctions is a possible wavefunction as long as it is normalized (the overall probability for finding the particle including all space sums to 1), continuous, and smooth (the first derivative of the wavefunction is continuous except where the potential energy is infinite).

|     | Q1  | Q6  | Q14 | Q16 | Q18 | Q19 | Q20 | Q24 | Q27 | Q30 | Q31 |
| --- | --- | --- | --- | --- | --- | --- | --- | --- | --- | --- | --- |
| (a) | 2%  | 3%  | 7%  | **40%** | 18% | 15% | 22% | 0%  | 7%  | 12% | **61%** |
| (b) | **40%** | 4%  | 17% | 5%  | 2%  | 5%  | 12% | 2%  | **43%** | 2%  | 14% |
| (c) | 5%  | 8%  | 9%  | 22% | 4%  | 20% | **45%** | **30%** | 29% | 17% | 4%  |
| (d) | 50% | **25%** | 19% | 28% | **38%** | 6%  | 7%  | 11% | 10% | 35% | 14% |
| (e) | 3%  | 58% | **46%** | 3%  | 37% | **51%** | 13% | 53% | 6%  | **29%** | 2%  |

Table II. Distribution of students' responses to questions related to the possible wavefunctions. Correct responses are in boldface.

The fact that a possible wavefunction need not be symmetric or anti-symmetric even if the potential energy has symmetry is tested in Question 1 (pictorial representation) and Question 30 (written representation). We placed these two questions far from each other in the survey to reduce the possibility that the students would refer to the picture in Question 1 while answering Question 30. In Question 1, 40% of the students selected the correct choice that all of the wavefunctions, including the asymmetric one, were possible for the given system. The most common difficulty, experienced by 50% of the students, was that the system did not allow for the asymmetric wavefunction. Question 30 was very challenging and only 29% of the students chose the correct response; 35% claimed that the possible wavefunctions for a particle in an even potential energy well must either be even or odd, and another 17% thought that the wavefunction must be symmetric but not necessarily about $x = 0$.

When the wavefunction was explicitly written as a linear superposition of the energy eigenstates, for example, $\psi(x) = \sum_n A_n \psi_n(x)$, many students recognized that this wavefunction is possible. In Question 6, over 90% of the students selected the correct choice (a) that $\Psi(x,0) = \sum_n A_n \psi_n(x)$ is a possible wavefunction for a particle in a 1D infinite square well, where $\psi_n(x)$ are the energy eigenfunctions. However, in Question 14, the wavefunction $A\sin^2(\pi x/a)$ is not expressed explicitly as a linear superposition of the energy eigenstates and more than 50% of the students mistakenly thought that it is not a possible wavefunction. Approximately 40% of the students chose the distracter choices (a) or (b), indicating that the possible wavefunction must satisfy the time independent Schrödinger equation. Another 9% incorrectly noted that $A\sin^2(\pi x/a)$ is a possible wavefunction for two particles and not a single particle. (A two particle wavefunction depends on two variables $x_1$ and $x_2$.)

Some students knew that the possible wavefunction must be continuous and smooth. However, they were unsure that any single valued, continuous, smooth, and normalized function satisfying the boundary conditions of the system is possible. In Question 16, a sketch of a wavefunction going to zero inside a finite square well was given. Students knew that for a finite square well, the particle has a nonzero probability of being in the classically forbidden region in a stationary state. However, they had the misconception that any possible wavefunction for this system must have a nonzero probability in the classically forbidden region. Only 40% of the students correctly noted that the wavefunction in Question 16 is possible.

A subgroup of the possible wavefunctions category is related to the bound and scattering states of a quantum system. When the energy of the quantum particle is less than the potential energy $V(x)$ at x = ± ∞, the particle is in a bound state. Otherwise, if the particle's energy is larger than $V(x)$ at x = ± ∞, it is in a scattering state. The bound states have a discrete energy spectrum and the scattering states have a continuous energy spectrum.

Questions 18 and 20 examine students' understanding of the shape of the bound/scattering state wavefunctions. The bound state wavefunctions go to zero at infinity so they can always be normalized. The scattering state wavefunctions are not normalizable because the probability of finding the particle is nonzero at infinity; a normalized wavefunction can be constructed using their linear superpositions. In Question 18, 20% of the students did not select statement (3), which suggests that they either thought that the scattering state wavefunctions are normalizable or they did not know that a linear superposition of the scattering state wavefunctions can be normalized. In individual interviews we found most students thought that scattering states could be normalized. Students who knew the general shape of the scattering state wavefunctions usually knew how to construct a normalized wave packet by taking the linear superposition of the scattering states. Also, in Question 18, 39% did not know that the scattering states have a continuous energy spectrum and claimed that energy is always discrete in quantum mechanics. In Question 20, students needed to understand that for a simple harmonic oscillator in its ground state, the probability of finding the particle is a maximum at the center, whereas classically the particle is more likely to be found close to the classical turning points. We found that 20% of the students who chose statement (3) in Question 20 thought that the quantum simple harmonic oscillator cannot be found in the region where $E < V(x)$. Discussions with individual students suggest that this difficulty often has its origin in their experiences with the turning points of a classical system. (In very few cases during the individual discussions did we find that this difficulty was due to experience with the quantum infinite well.) 22% of the students who selected choice (a) did not know that the first excited state wavefunction of the simple harmonic oscillator is zero at $x = 0$ in the middle of the potential energy well; 19% of the students who chose (b) or (d) did not realize that for a very high energy stationary state, the probability distribution for finding the particle is consistent with the classical distribution according to Bohr's correspondence principle; the ground state of a quantum system can have very different behavior from the classical behavior.

Questions 24 and 27 ask that students decide whether a given potential energy $V(x)$ allows for bound states or scattering states. Question 24 uses a pictorial representation showing four different potential energy wells. The distracter that the students found challenging was picture (3) in which the potential energy of the well bottom was greater than the potential energy at infinity (which is zero). Therefore, no bound state can exist in this potential energy well. About 2/3 of the students failed to notice the difference between pictures (3) and (4). They had the misconception that any potential energy $V(x)$ that has the shape of a "well" would allow for bound states if there were classical turning points. In Question 24, 85% of the students had selected picture (2) as the potential energy that allows both bound and scattering states. Question 27 asked students to choose the Hamiltonian operators that have only a discrete energy spectrum from three choices. The most common mistake, by 40% of the students, was that the finite square well allows only discrete energies. There are at least two possible sources for students' difficulties in Question 27: they might have difficulty constructing the correct pictorial representation from the mathematical representation, or fail to recognize the connection between the bound/scattering states and the discrete/continuous energy spectrum.

Questions 19 and 31 focus on the misconception that a given particle may be in a bound or a scattering state depending on its location. This notion often has its origin in students' classical experience. In Question 19, 15% mistakenly thought that the particle could have different energies in different regions. In fact, if a quantum particle is in an energy eigenstate, it has a definite energy and does not have different energies in different regions. If the particle is not in an energy eigenstate, it does not have a definite energy until a measurement of its energy is performed. In Question 19, 20% of the students selected incorrect option (c), and 6% selected incorrect option (d). Individual discussions suggest that students who selected option (c) often incorrectly thought that the particle is in a bound state when it is in the classically allowed region and is in a scattering state when it is in a classically forbidden region. A similar difficulty was found in Question 31. In particular, 14% of the students selected incorrect option (b) and claimed that statement (3) is correct, which indicates that the students did not realize that whether a state is a bound or a scattering state depends only on the energy of the particle compared to the potential energy at ± infinity.

**B. Expectation Values**

Questions 2 and 23 ask students to evaluate the time dependence of the expectation values of different physical observables in a stationary or a non-stationary state respectively. In Question 2 the initial state is an energy eigenstate, so the expectation value of any time-independent operator is time-independent. The most common mistake in Question 2 was the belief that the expectation values of the position and momentum operators depend on time in a stationary state. The initial state in Question 23 is a linear superposition of the energy eigenstates $\frac{1}{\sqrt{2}}(\psi_1 + \psi_2)$, which is not a stationary state. The expectation value of the energy is time independent because the probability of obtaining energies $E_1$ or $E_2$ is always 50%, but the expectation value of the position $\langle \hat{x} \rangle$ depends on time. Students need not evaluate the integrals to determine the correct response if they realize that for a non-stationary state, the probability density changes with time. Also the position and momentum operators do not commute with the Hamiltonian so their expectation values depend on time in a non-stationary state. 13% of the students mistakenly thought that all the expectation values of the position, momentum, and energy depend on time when the wavefunction is not a stationary state. 15% chose option (c) (only the expectation value of the energy depends on time), which is the opposite of the correct option (d). In contrast, only 5% of the students in Question 2 thought that the expectation value $\langle \hat{H} \rangle$ depends on time, but the expectation values $\langle \hat{x} \rangle$ or $\langle \hat{p} \rangle$ do not when the system is in a stationary state.

Question 12 asks students to compare the expectation values of different physical observables at time $t$ for an infinite square well for the initial states $\frac{1}{\sqrt{2}}(\psi_1 + \psi_2)$ and $\frac{1}{\sqrt{2}}(\psi_1 + i\psi_2)$, which are different linear combinations of the same energy eigenstates. The expectation values of the energy for the two initial states are the same. Because the relative phases of $\psi_1$ and $\psi_2$ are different for the two states, the shape of the probability density is different at time $t$. Therefore, the expectation values of the position (or momentum) of the particles are not the same in the two states. Only 29% of the students chose the correct response. 28% thought that the relative phases would not affect the expectation values of position and momentum. Another 27% incorrectly thought that the expectation value of energy would also be affected by the relative phase. Similar to Question 23, 14% thought that the superposition of energy eigenstates with different relative phases would give different expectation values of the energy, but the expectation value of position or momentum would not change.

Question 9 investigates whether the students understand different ways to represent the expectation value of the energy. The expectation value is the average of a large number of measurements on identically prepared systems and is equal to the sum of the possible values multiplied by their probabilities. It can also be written as $\langle E \rangle = \int_0^a \psi^*(x,0) \hat{H} \psi(x,0) dx$. 21% of the students incorrectly thought that $\langle E \rangle = \frac{1}{3} E_1 - \frac{2}{3} E_2$ (incorrect sign) and 18% thought that only the integral form $\langle E \rangle = \int_0^a \psi^*(x,0) \hat{H} \psi(x,0) dx$ is correct. They did not connect the definition of the expectation value with its physical meaning, which is the average of a large number of measurements on identically prepared systems. In Question 10 the initial state is the same as in Question 9, but students need to evaluate the expectation value at time $t > 0$. 74% of the students selected the correct answer to Question 10. However, many might not understand that the expectation value of energy is time-independent. In particular, students who answered Question 9 incorrectly might answer Question 10 correctly because only one of the choices (algorithmic method for calculating the expectation value) is correct. In the future versions of the survey, we plan to use $\langle E \rangle = \frac{1}{3} E_1 + \frac{2}{3} E_2$ as one of the correct choices in Question 10.

Question 25 involves the degeneracy in a 1D free particle system. The stationary state wavefunctions $e^{ikx}$ and $e^{-ikx}$ have momentum in the opposite directions, but have the same energy, and their superposition $e^{ikx} + e^{-ikx}$ is an energy eigenstate. The expectation value of momentum is zero, but that of the energy is nonzero. 23% of the students did not know that $e^{ikx} + e^{-ikx}$ is a stationary state. Also, 27% of the students incorrectly selected the choice (a). They knew that $e^{ikx} + e^{-ikx}$ is a stationary state, but did not realize that $e^{ikx}$ is a momentum eigenstate with a definite value of momentum, and the expectation value of momentum is zero in the state $e^{ikx} + e^{-ikx}$.

|     | Q2  | Q9  | Q10 | Q12 | Q23 | Q25 | Q26 |
|-----|-----|-----|-----|-----|-----|-----|-----|
| (a) | 4%  | 3%  | 7%  | 11% | 6%  | 27% | **45%** |
| (b) | 8%  | 4%  | 1%  | 3%  | 10% | 21% | 2%  |
| (c) | 5%  | 18% | 6%  | **29%** | 15% | 2%  | 25% |
| (d) | 14% | 18% | **74%** | 28% | **50%** | **38%** | 1%  |
| (e) | **69%** | **56%** | 11% | 27% | 13% | 5%  | 23% |

Table III. Distribution of students' responses for questions related to expectation values.

**C. Stationary State**

Questions 3 and 4 require students to decide whether the initial state, $\Psi(x,0)$, is a stationary state before they calculate the probability density $|\Psi(x,t)|^2$ at time $t$. In response to Question 3, 78% of the students knew that $\sqrt{2/a}\sin(5\pi x/a)$ is an energy eigenstate with energy $E_5$, so the probability density $|\Psi(x,t)|^2$ is time-independent, but 18% failed to multiply the complex conjugate correctly when they calculated the probability density so their responses had the incorrect phase factor $\exp(-i2E_5 t/\hbar)$. In Question 4, only 35% realized that $A\sin^5(\pi x/a)$ is not a stationary state but a linear superposition of different stationary states. In particular, 49% mistakenly thought that the probability density in Question 4 is time independent, similar to Question 3.

When the potential energy of a quantum system is changed suddenly, a stationary state of the old system might not be a stationary state of the new system. When the infinite square well was expanded suddenly at time $t = 0$ in Question 15 the ground state at time $t < 0$ is not a stationary state at time $t > 0$. Only 42% of the students correctly noted that the probability density function evolves in time for all $t > 0$. The most common misconception was that the old ground state would eventually evolve into a new stationary state. 26% of the students thought that the wavefunction would evolve into the new ground state, and 19% thought that the system would evolve into the new first excited state because the ground state wavefunction of the old system is similar in form to the first excited state of the new system for $0 \leq x \leq a$. However, because the initial wavefunction is zero in the region $a < x \leq 2a$, the old ground state is a linear superposition of the stationary states of the new system after the well has expanded. The students did not realize that if the initial state is not a stationary state of the new system, the time evolution would not cause the wavefunction to evolve into a stationary state of the new system.

Question 28 assesses whether students can distinguish between the stationary states and the eigenstates of other physical observables. The most common misconception was that an eigenstate of a physical observable is a stationary state. In particular, half of the students incorrectly thought that statement (1) in Question 28, which states that the stationary states refer to the eigenstates of any operator corresponding to any physical observable, is correct. Another 10% did not choose statement (1), but incorrectly claimed that if the particle has a well-defined position in the initial state, the position of the particle is well defined for all future times.

|     | Q3  | Q4  | Q6  | Q15 | Q25 | Q28 | Q29 |
|-----|-----|-----|-----|-----|-----|-----|-----|
| (a) | 2%  | 2%  | 3%  | 6%  | 27% | 13% | 7%  |
| (b) | 18% | 13% | 4%  | 26% | 21% | **36%** | 15% |
| (c) | 0%  | 0%  | 8%  | 19% | 2%  | 25% | **43%** |
| (d) | **78%** | 49% | **25%** | **42%** | **38%** | 10% | 14% |
| (e) | 1%  | **35%** | 58% | 6%  | 5%  | 12% | 18% |

Table IV. Distribution of students' responses for questions related to the stationary states versus non-stationary states.

## D. The Role of the Hamiltonian

The Hamiltonian governs the time evolution of the system according to the time dependent Schrödinger equation. In Question 29 students were asked about the role of the Hamiltonian in a quantum system. The most common misconception was that the Hamiltonian determines the shape of a position eigenfunction. 15% of the students did not know that the Hamiltonian governs the time evolution. Another 7% did not relate the Hamiltonian to the shape of the stationary state wavefunctions. Individual discussions suggest that sometimes this mistake originates from their misunderstanding of a stationary state as an eigenstate of any operator corresponding to a physical observable. Students' response to Question 26 suggests that most knew that the Hamiltonian is the sum of the potential energy and kinetic energy, but their response to Question 27 suggests that more than half of them had difficulty selecting the Hamiltonian operators that have only a discrete energy spectrum.

|     | Q26 | Q27 | Q29 |
|-----|-----|-----|-----|
| (a) | **45%** | 7%  | 7%  |
| (b) | 2%  | **43%** | 15% |
| (c) | 25% | 29% | **43%** |
| (d) | 1%  | 10% | 14% |
| (e) | 23% | 6%  | 18% |

Table V. Distribution of students' responses for questions related to the Hamiltonian.

## E. Time dependence of the wavefunction

The stationary state wavefunction at time $t$ satisfies both the time independent and time dependent Schrödinger equations. However, a linear superposition of the stationary states does not have a definite value of energy even at $t = 0$, for example, $\hat{H}(\psi_1 + \psi_2) = E_1\psi_1 + E_2\psi_2 \neq E(\psi_1 + \psi_2)$. In Question 6 about 70% of the students incorrectly thought that the superposition state $\psi(x) = \sum_n A_n \psi_n(x)$ is an energy eigenstate which satisfies the time independent Schrödinger equation. Only 25% selected the correct answer that $\psi(x) = \sum_n A_n \psi_n(x)$ is not the solution of the time independent Schrödinger equation, but its time evolution $\psi(x,t) = \sum_n A_n \psi_n(x) \exp(-iE_n t/\hbar)$ satisfies the time dependent Schrödinger equation. Further interviews indicate that many undergraduate and graduate students hold the misconception that the time independent Schrödinger equation is satisfied for any possible wavefunction.

Question 17 tests the understanding of the time dependence of a position eigenfunction. The position eigenfunction is a delta function, which can be written as a linear superposition of energy eigenfunctions. The position eigenfunction is not a stationary state wavefunction and changes with time. 44% of the students selected the correct statement (3) [in options (c) and (e)], but some of them [who chose option (c)] did not answer the question correctly because they did not know that the wavefunction would become peaked after a position measurement. 39% of the students selecting statement (2) held the misconception that a position eigenfunction would evolve with time after the measurement, but eventually return to the state right before the position measurement was performed.

|     | Q2  | Q3  | Q4  | Q6  | Q15 | Q17 | Q22 | Q29 |
|-----|-----|-----|-----|-----|-----|-----|-----|-----|
| (a) | 4%  | 2%  | 2%  | 3%  | 6%  | 15% | 5%  | 7%  |
| (b) | 8%  | 18% | 13% | 4%  | 26% | 7%  | 3%  | 15% |
| (c) | 5%  | 0%  | 0%  | 8%  | 19% | 5%  | **70%** | 43% |
| (d) | 14% | **78%** | 49% | **25%** | 42% | 32% | 14% | 14% |
| (e) | **69%** | 1%  | **35%** | 58% | 6%  | **39%** | 4%  | 18% |

Table VI. Distribution of students' responses for questions related to the time dependence of the wavefunction.

## F. Measurements

When calculating the probability of obtaining a certain value in a measurement of a physical observable, students often incorrectly think that the operator corresponding to the observable must be explicitly involved in the expression. For example, in Question 5, 30% chose the distractor $\int_{x}^{x+dx} x|\psi_1(x)|^2 dx$ as the probability of finding the particle in the region between $x$ and $x+dx$. They did not realize that $|\psi_1(x)|^2 dx$ is the probability density of finding the particle between $x$ and $x + dx$. In Question 11, 33% incorrectly thought that $\left|\int_0^a \psi_n^*(x)\hat{H}\Psi(x,0)dx\right|^2$ is the probability of measuring the energy $E_n$ at time $t = 0$ instead of the correct expression $\left|\int_0^a \psi_n^*(x)\Psi(x,0)dx\right|^2$. Students often did not realize that the required information about the energy measurement is obtained by projecting the state of the system along the energy eigenstate (multiplying the wavefunction by $\psi_n^*(x)$ before integrating). Further interviews indicate that students held a common misconception that the Hamiltonian acting on a state represents an energy measurement. This incorrect notion is an overgeneralization of the fact that the system is in a stationary state after the energy measurement.

Questions 7 and 8 investigate students' understanding of the energy measurement outcomes for the superposition state $\sqrt{2/7}\psi_1(x)+\sqrt{5/7}\psi_2(x)$. The only possible energies are the ground state energy $E_1$ and the first excited state energy $E_2$. When the energy $E_2$ is obtained, the wavefunction collapses to $\psi_2(x)$. In Question 7, 32% incorrectly claimed that the wavefunction would collapse first but eventually return to the initial state $\sqrt{2/7}\psi_1(x)+\sqrt{5/7}\psi_2(x)$. Another 13% did not note that the wavefunction would collapse and thought that the system will remain in the initial state even after the measurement. In Question 8, 20% claimed they could measure not only $E_1$ and $E_2$, but any possible energy $E_n$ ($n$ is a positive integer), and 25% claimed that the probabilities for measuring any energy $E_n$ are equal.

Question 13 examines students' understanding of consecutive quantum measurements, for example, measuring the energy of a system immediately after a position measurement. For a 1D infinite square well with the initial state $\frac{1}{\sqrt{2}}(\psi_1+\psi_2)$, a position measurement will collapse the wavefunction to a delta function which is a superposition of many energy eigenfunctions. So we can obtain a higher order energy $E_n$ ($n > 2$) for the energy measurement of the system after the position measurement. Only 31% of the students correctly answered Question 13 and realized that the state of the system changed after the position measurement. 40% mistakenly thought that the result could be only energy $E_1$ or $E_2$, which corresponds to the initial state before the position measurement.

Question 22 asks students to predict an unknown quantum state for a simple harmonic oscillator in a linear superposition of the ground and third excited states by a measurement. When there is a large ensemble of particles in the state $a\psi_0 + b\psi_3$ (*a* and *b* are coefficients whose absolute square is to be determined), we can measure the energy of each particle and count the number of particles collapsing to the states $\psi_0$ and $\psi_3$, and then calculate the proportions of $\psi_0$ and $\psi_3$ to estimate the absolute squares of *a* and *b*. 70% of the students knew that the measurement would change the state of the particle so they had to prepare the particle in the initial state again before making another measurement. 17% of the students mistakenly thought that the wavefunction would automatically return to the original state a long time after the measurement. The other students who selected statement (1) in Question 22 did not realize that the wavefunction changes after the energy measurement.

|     | Q5  | Q7  | Q8  | Q11 | Q13 | Q17 | Q20 | Q22 | Q28 |
|-----|-----|-----|-----|-----|-----|-----|-----|-----|-----|
| (a) | **44%** | **45%** | 3%  | 33% | 40% | 15% | 22% | 5%  | 13% |
| (b) | 2%  | 8%  | **74%** | 43% | *31%* | 7%  | 12% | 3%  | **36%** |
| (c) | 30% | 13% | 15% | 12% | 6%  | 5%  | **45%** | **70%** | 25% |
| (d) | 4%  | 18% | 5%  | 8%  | 11% | 32% | 7%  | 14% | 10% |
| (e) | 19% | 14% | 1%  | 3%  | 9%  | *39%* | 13% | 4%  | 12% |

Table VII. Distribution of students' responses for questions related to quantum measurement.

**G. Other**

The position-momentum uncertainty principle is a central principle of quantum mechanics. Written responses and individual discussions suggest that students are often unclear about the difference between the quantum uncertainty principle and experimental uncertainty. Students often have the misconception that the uncertainty in position or momentum is about the expectation value of the position or momentum of the particle. In Question 21, 22% of the students who selected statement (1) incorrectly claimed that the uncertainty in position is smaller when the expectation value of the momentum is larger. About 23% of the students who selected options (d) or (e) claimed that the expectation value of the position is larger when the expectation value of the momentum is smaller, that is, $\langle x \rangle \langle p \rangle \geq \text{constant}$. The students were unclear that the uncertainty of a physical observable depends on the standard deviation, instead of the expectation value of that observable for a given wavefunction.

Question 26 is related to the Ehrenfest theorem. In the Schrödinger formalism the expectation values obey the classical laws of motion. To determine the time-dependence, many students substituted the classical variables by the quantum operators instead of the expectation value. For example, 50% of the students who selected statement (1) incorrectly claimed that the momentum operator $\hat{p}$ is equal to $m \frac{d\hat{x}}{dt}$ and about 26% also mistakenly claimed that $\frac{d\hat{p}}{dt} = -\frac{\partial V(\hat{x})}{\partial x}$. It is important to help students build a robust knowledge structure so that they do not incorrectly over-generalize their experiences from classical physics.

|     | Q21 | Q25 | Q26 |
| --- | --- | --- | --- |
| (a) | **59%** | 27% | **45%** |
| (b) | 5% | 21% | 2% |
| (c) | 7% | 2% | 25% |
| (d) | 6% | **38%** | 1% |
| (e) | 17% | 5% | 23% |

Table VIII. Distribution of students' responses for questions related to other concepts.

## V. SUMMARY

Identification of students' difficulties can help the design of better instructional strategies and learning tools to improve students' understanding. We have developed a research-based multiple choice survey to assess students' conceptual understanding of quantum mechanics in one spatial dimension. The alternative choices for the multiple-choice questions on the survey often deal with the common difficulties found in these investigations.

We found that the advanced undergraduate and graduate students have many common difficulties and misconceptions about various topics. We also investigated the extent to which research-based learning tools[17,22] can help students learn these concepts and found that the difficulties were significantly reduced when students used concept tests and Quantum Interactive Learning Tutorials. Students who used research-based learning tools in their quantum mechanics courses not only performed better on the survey when it was administered at the end of the same semester in which the relevant concepts were covered but performed equally well after an entire semester suggesting good retention of the concepts. The survey can be administered to students in upper-level undergraduate courses after instruction. It can also be used as a preliminary test for graduate students to evaluate their background knowledge in quantum mechanics before they take graduate-level quantum mechanics courses. Those developing instructional strategies to improve student understanding of quantum mechanics can take into account the difficulties that were brought out by the survey.

ACKNOWLEDGEMENTS

We are very grateful to all the faculty who reviewed the various components of the survey at several stages and provided invaluable feedback. We are also very thankful to all the faculty who administered the test. This work is supported by the National Science Foundation.